# Representability problems for coarse-grained water potentials

Margaret E. Johnson[1], Teresa Head-Gordon[1,2,*] and Ard A. Louis[3,*]

[1]UCSF/UCB Joint Graduate Group in Bioengineering and [2]Department of Bioengineering, University of California, Berkeley, CA 94720

[3]Dept. of Chemistry, University of Cambridge, Lensfield Road, CB2 1EW, Cambridge, UK

The use of an effective intermolecular potential often involves a compromise between more accurate, complex functional forms and more tractable simple representations. To study this choice in detail, we systematically derive coarse-grained isotropic pair potentials that accurately reproduce the oxygen-oxygen radial distribution function of the TIP4P-Ew water model at state points over density ranges from 0.88-1.30g/cc and temperature ranges from 235K-310K. Although by construction these effective potentials correctly represent the isothermal compressibility of TIP4P-Ew water, they do not accurately resolve other thermodynamic properties such as the virial pressure, the internal energy or thermodynamic anomalies. Because at a given state point the pair potential that reproduces the pair structure is unique, we have therefore explicitly demonstrated that it is impossible to simultaneously represent the pair-structure and several key equilibrium thermodynamic properties of water with state-point dependent radially symmetric pair potentials. We argue that such *representability problems* are related to, but different from, more widely acknowledged *transferability problems*, and discuss in detail the implications this has for the modeling of water and other liquids by coarse-grained potentials. Nevertheless, regardless of thermodynamic inconsistencies, the state-point dependent effective potentials for water do generate structural and dynamical anomalies.

*Corresponding authors

## I. Introduction

The properties of a molecular fluid are fundamentally determined by the Schrödinger equation for the nuclei and the electrons, but for most practical applications such formulations are far too onerous to solve. To make progress, a series of coarse-graining approximations must be made, with each step leading to a less complete, but more tractable representation of the fluid. One popular route integrates out the quantal degrees of freedom to represent the interactions between molecular species by classical effective potentials. These can in turn be determined by analytic considerations [1], or be extracted by fitting to thermodynamic and structural properties of experiment or high quality calculations on model systems. Making a judicious choice between the accuracy of more complex representations versus a practical utility of simpler models comprises the art of asking the right questions of a scientific problem.

At their best, coarse-graining strategies facilitate insight into the basic underlying physics, stripped of all non-essentials. For example, a fully quantum mechanical calculation of the properties of simple liquids like Argon could obscure the important observation that their structure is dominated by the hard repulsions of the effective inter-atomic interactions[2,3]. Nevertheless, it also seems intuitively obvious that the simpler the representation of a molecular fluid, the less faithful the resolution of the complete suite of underlying physical properties. In other words, we expect that in science, as in life, there is no such thing as a free lunch.

In the present paper we carefully examine the coarse-graining of the multi-site TIP4P-Ew water model[4] to a series of simpler isotropic single-site potentials [5,6] by an inversion of the oxygen-oxygen radial distribution function at each characterized state point in temperature and density [2,7,8]. We focus on water because many of its well-characterized structural[9] and thermodynamic[6] anomalies are postulated to be related to its strong orientational order [9]. One might therefore expect that integrating out these orientational degrees of freedom to derive an isotropic pair potential would be a worst-case scenario, and a useful foil to highlight coarse-graining pitfalls.

It was first shown by Henderson, that, under fairly general conditions, an isotropic potential derived to reproduce the pair structure of a fluid is unique up to a constant [10-12]. Although any given state point, the pair correlations generated through this route are, by construction, correctly reproduced, we find that for our water model, higher order structural correlations, as well as thermodynamic properties like the virial pressure and the internal energy, are poorly rendered. The Henderson uniqueness proof therefore implies a *representability problem* as a general feature of effective potentials, i.e. they cannot simultaneously resolve all the properties of the more complex reference system at a given state point. In fact we have suggested that these representability problems are a more general feature of effective potentials [13].

In this work we also pose a different question. Are representability problems as severe when, rather than comparing quantitative properties, we study *qualitative trends* for a family of effective potentials derived from many T,ρ state points? Can they reproduce the subtle anomalous water *trends* in structure, thermodynamics, and dynamics exhibited by the more complicated potential? We find that the family of effective potentials does generate anomalous trends in structure (decreasing order with compression) when considered under a general orientational order metric, and anomalous transport properties (exhibiting minima and maxima in the self diffusion coefficient) as a function of density at several low temperature isotherms. By contrast, we do not find thermodynamic anomalies (density maxima) when comparing results across state-points.

We proceed as follows: Section II defines and contrasts the concepts of *transferability* and *representability* for effective potentials, following the analysis of one of us[13], where this was done for soft-matter systems. We also work out under what conditions an effective potential is *unique*. Section III describes reference molecular dynamics (MD) calculations for the TIP4P-Ew water model[4] and shows how, for a given state-point, the oxygen-oxygen radial distribution function $g_{OO}(r)$ of TIP4P-Ew water can be reliably inverted through an empirical potential structural refinement (EPSR) procedure[8] to obtain a unique effective isotropic pair potential $v_g(r)$. Section IV compares the resultant thermodynamic, structural, and diffusive behavior produced by the reference and coarse-grained models. Finally, in Section V, we conclude, making several recommendations for the coarse-graining of substances like water to isotropic pair potentials.

## II. Transferability, representability and uniqueness

### A. Hamiltonians with pair and triplet interactions.

Consider a one-component reference fluid interacting with a three-body Hamiltonian [14,15] of the form:

$$H = K + \sum_{i<j} w^{(2)}(r_{ij}) + \sum_{i<j<k} w^{(3)}(\mathbf{r}_{ij}, \mathbf{r}_{jk}, \mathbf{r}_{ki}) \tag{1}$$

where $\mathbf{r}_i$ denotes the position of particle $i$ and $\mathbf{r}_{ij} = \mathbf{r}_i - \mathbf{r}_j$ and $r_{ij} = |\mathbf{r}_i - \mathbf{r}_j|$. $K$ is the kinetic energy operator, $w^{(2)}(r)$ is an isotropic pairwise additive potential, and $w^{(3)}(r_{ij}, r_{jk}, r_{ki})$ is a triplet or three-body potential. A common example of the latter would be the Axilrod-Teller triple dipole form [16]. Three-body potentials are expensive and cumbersome to simulate, and so it is often desirable to coarse-grain them to a simpler isotropic representation.

*Transferability problems.* One way to derive a simpler pairwise potential would be to calculate the excess internal energy $U$ at a state point (N,V,T), which for a homogeneous fluid is given by

$$U(N,V,T) = \frac{1}{2}\rho^2 \int d\mathbf{r}_1 d\mathbf{r}_2 g(r_{12}) w^{(2)}(r_{12}) + \frac{1}{6}\rho^3 \int d\mathbf{r}_1 d\mathbf{r}_2 d\mathbf{r}_3 g^{(3)}(\mathbf{r}_{12}, \mathbf{r}_{23}, \mathbf{r}_{31}) w^{(3)}(\mathbf{r}_{12}, \mathbf{r}_{23}, \mathbf{r}_{31}) \quad (2)$$

where $g(r)$ and $g^{(3)}(\mathbf{r}_{12}, \mathbf{r}_{23}, \mathbf{r}_{31})$ are the homogeneous pair and triplet radial distribution functions at the density $\rho$. For a given state point, a new isotropic effective pair potential $v_U^{\text{eff}}(r)$ can be defined such that the excess internal energy of the reference fluid, given by Eq. (2), is exactly reproduced by the standard formula for a two-body Hamiltonian:

$$U(N,V,T) = \frac{1}{2}\rho^2 \int d\mathbf{r}_1 d\mathbf{r}_2 g(r_{12}) v_U^{\text{eff}}(r_{12}) \quad (3)$$

Comparing Eqns. (2) and (3) and rewriting the potential as

$$v_U^{\text{eff}}(r) = w^{(2)}(r) + \delta v_U(r) \quad (4)$$

results in

$$\delta v_U(r) = \frac{1}{3}\rho \int d\mathbf{r}_3 \frac{g^{(3)}(\mathbf{r}_{12}, \mathbf{r}_{23}, \mathbf{r}_{31})}{g(r_{12})} w^{(3)}(\mathbf{r}_{12}, \mathbf{r}_{23}, \mathbf{r}_{31}) \quad (5)$$

where it has been assumed that the two-body radial-distribution function, $g(r)$, is unchanged from the original reference system. The procedure above only fixes a single number, $U(N,V,T)$, and so there is some flexibility in the exact form of the potential since any function which integrates to zero in Eq.(3) could be always be added to $v_U^{\text{eff}}(r)$. Nevertheless, Eq.(5) is useful for illustrative purposes because it can be written in a simple closed form. Note that this derivation, which we will call the energy coarse-graining route, can easily be generalized to higher order many-body terms in Eq.(1). It is similar in spirit to routes often used in other contexts.

Since $g(r)$ and $g^{(3)}(\mathbf{r}_{12}, \mathbf{r}_{23}, \mathbf{r}_{31})$ depend on the intensive state variables $\rho$ and $T$, $\delta v_U(r)$ will itself also vary with state point. This property leads to what are often called *transferability problems*, where an effective potential derived in one context does not perform well in a different context (i.e. another state point). Such problems are very well recognized in the literature. For example protein force fields are derived (in part) from fitting to the conformational energies of small peptide fragments that hopefully are transferable to the longer protein chain. In the case of water, non-polarizable force fields attempt to reproduce many ambient state point properties, and more recently, the temperature of maximum density at 1atm, but require validation over other state points for which they are not explicitly parameterized.

It is clear from Eq.(5) that the larger the three-body interaction, the more important the correction $\delta v_U(r)$. Moreover,

$$\lim_{\rho \to 0} \delta v_U(r) = 0, \quad (6)$$

reflecting the fact that the relative influence of three-body (and if present, higher order many-body) interactions becomes less important with decreasing density. This argument can, however, be subtle. For example at small, but finite density, water reaches a limit of tensile strength, below which it is mechanically unstable. [17]

*Representability problems.* Instead of fixing the potential to reproduce the internal energy, one could instead require it to reproduce the pair structure. We will call this the pair structure coarse-graining route. There is no known exact closed form expression like Eq. (5), but an expansion in density yields [15,18,19]:

$$v_g^{eff}(r) = w^{(2)}(r) + \delta v_g(r) \tag{7}$$

with

$$\beta \delta v_g(r) = \rho \int d\mathbf{r}_3 \left[1 - \exp\left(-\beta w^{(3)}(\mathbf{r}_{12}, \mathbf{r}_{23}, \mathbf{r}_{31})\right)\right] g(r_{23}) g(r_{31}) \tag{8}$$

and $\beta = (k_B T)^{-1}$ where $k_B$ is Boltzmann's constant. Clearly this correction does not have the same form as Eq.(5) (although it has the same $\rho \to 0$ limit). To lowest order in $\rho$ and $w^{(3)}$, the ratio between the two corrections is

$$\frac{\delta v_U(r)}{\delta v_g(r)} = \frac{1}{3} + O\left[\left(w^{(3)}\right)^2; \rho^2\right] \tag{9}$$

In other words the energy and pair structure coarse-graining routes yield corrections to the bare pair-potential which may differ by as much as a factor three! As the density is increased, this difference might decrease or increase, but clearly corrections to the bare pair potential $w^{(2)}(r)$ that take into account the effect of a three-body potential depend on which physical property one is attempting to reproduce. In an important study of liquid Argon, van der Hoef and Madden[20] used a similar analysis to that above, showing that if in addition one fits a potential to reproduce the virial pressure, then the ensuing correction to the bare pair potential is different from both $\delta v_g^{eff}(r)$ and $\delta v_U(r)$. Problems of this type have been called *representability problems* [13]: For a given state point, it is not possible to simultaneously represent multiple physical properties of a system with a single coarse-grained potential. This contrasts with transferability problems that relate potentials at different state points.

**B. Uniqueness of effective potential**

One might argue that the representability problems described above simply reflect the particular choices of coarse-graining method -- a lack of imagination as it were – and that there still exists a hypothetical effective isotropic pair potential $v^{eff}(r)$ that will simultaneously represent the pair structure $g(r)$ and a thermodynamic property like the excess internal energy. As shown above, it is indeed true that $v_U(r)$ is not unique, since there may be a whole family of potentials that all reproduce $U(N,V,T)$ at a given state point. By contrast, we will argue that for any given state-point, $v_g^{eff}(r)$ is unique.

Henderson[10] first showed, using arguments very similar to those used by Hohenberg and Kohn[21] in their famous proof relating the one-body potential to the one-body density (which laid the foundation for density functional theory), that "the pair potential $v(r)$ which gives rise to a radial distribution function $g(r)$ is unique up to a constant." An extended proof for orientational correlations can be found in a book by Gray and Gubbins[11], while a more rigorous mathematical discussion is provided by Chayes and Chayes[12].

While these studies do not prove that, given a *g(r)*, there always exists a pairwise *v(r)* that can generate the same pair correlations, they do show that if such a potential can be found, then it will be unique, up to a trivial constant. The argument goes as follows: if a pairwise potential $v_g^{eff}(r)$ can be found that reproduces a given *g(r)* at a given state point, then, irrespective of what Hamiltonian originally generated *g(r)*, this potential can be used to define a new (fictitious) pairwise Hamiltonian which generates *g(r)* at that state point, and which moreover satisfies the conditions under which the uniqueness theorem was derived[10-12]. Note that there will be a separate fictitious Hamiltonian at each state point, but this doesn't affect the uniqueness proof.

This theorem does not imply that higher order correlation functions such as $g^{(3)}(\mathbf{r}_{12}, \mathbf{r}_{23}, \mathbf{r}_{31})$ are correctly reproduced[22]. Of course if *g(r)* is initially generated by a Hamiltonian with only an isotropic pairwise potential energy term, then a correct inversion to $v_g^{eff}(r)$ will exactly reproduce this potential due to the uniqueness theorem, and therefore also correctly generate the higher order correlation functions. But if *g(r)* is produced by a Hamiltonian with anisotropic potential terms, e.g. Eq. (1), then $v_g^{eff}(r)$ is expected to generate different three-body and higher order correlations. For an example from soft-matter where this difference is demonstrated explicitly see reference[23].

Plugging the unique $v_g^{eff}(r)$ into the two-body formula, Eq. (3) in order to extract the excess internal energy U(N,V,T) will generally not reproduce the correct internal energy of the underlying reference system [13,20]. So even though $v_U(r)$ itself can have several different forms, these are not expected to be the same as $v_g^{eff}(r)$, ensuring that the representability problems cannot be easily evaded.

Another point to keep in mind is that Eq. (3) assumes that *g(r)* is identical to that of the reference system. What is normally done in practice is that $v_U(r)$ is also used to generate the radial distribution function, which we shall call *g_U(r)*. By the uniqueness theorem, *g_U(r)* ≠ g(r). Nevertheless, the fact that $v_U(r)$ itself is only specified up to a function that integrates to zero in Eq.(3) suggests that although the exact *g(r)* may not be attainable, if one chooses to exactly reproduce U(N,V,T), the best strategy may be to exploit the non-uniqueness of $v_U(r)$ to simultaneously minimize the difference between *g_U(r)* and *g(r)*.

An advantage of the unique $v_g^{eff}(r)$ is that, since by construction it generates the correct pair-correlations, thermodynamic properties can be extracted through the compressibility route:

$$\rho k_B T \chi_T = 1 + 4\pi\rho \int r^2 [g(r)-1] dr, \qquad (10)$$

Where $|_T$ is the isothermal compressibility at temperature T and density ρ. This relationship can be derived in the grand canonical ensemble and is independent of the form of the underlying Hamiltonian[2]. However, its practical use is limited because one needs *g(r)* to derive $v_g^{eff}(r)$ in the first place, and moreover many thermodynamic properties of interest (like the pressure) are related to the compressibility by an integral over density, which means deriving a new potential at each new state point. (See however [24] for an example where this strategy leads to important speedups for a two-component system.)

## C. Hamiltonians with pairwise multi-site interactions

In this paper we are focusing on the coarse-graining of a multi-site pairwise representation of water to an effective description based on a one-site isotropic potential. There are many similarities, but also some differences with the analysis carried out above for three-body Hamiltonians like *eqn.* (1).

The internal energy of a one-component fluid of particles interacting with a pairwise, but multi-site potential, can be written as[2]:

$$U(N,V,T) = 2\pi \frac{N^2}{V} \sum_{\alpha,\beta} \int v_{\alpha\beta}(r) g_{\alpha\beta}(r) r^2 dr \qquad (11)$$

where the $g_{\alpha\beta}(r)$ are the inter-site radial distribution functions and the $v_{\alpha\beta}(r)$ are the site-site potentials for sites $\alpha, \beta$ on different molecules. An effective representation based on isotropic potentials could be derived through the energy coarse-graining route by picking a particular site-site radial distribution function, which we shall call $g_{AB}(r)$, and insisting that the internal energy is reproduced by the simple two-body formula:

$$U(N,V,T) = 2\pi \frac{N^2}{V} \int_0^\infty g_{AB}(r) v_U^{eff}(r) r^2 dr \qquad (12)$$

which, in direct analogy with Eq. (3), defines an effective potential $v_U^{eff}(r)$. An analytic form for $v_U^{eff}(r)$ is easily obtained by comparing Eq. (11) with Eq. (12):

$$v_U^{eff} = v_{AB}(r) + \delta v_U(r) \qquad (13)$$

where

$$\delta v_U(r) = \frac{\sum'_{\alpha,\beta} g_{\alpha\beta}(r) v_{\alpha\beta}(r)}{g_{AB}(r)} \qquad (14)$$

and the ' on the sum means that the term with $\alpha\beta=AB$ is left out.

Just as was found for the many-body Hamiltonians, this energy coarse-graining route yields an effective potential that is state dependent, since it is mediated by the correlation functions $g_{\alpha\beta}(r)$. But in contrast to case for a many-body Hamiltonian described by Eq. (6), the low-density limit is not normally zero because it includes an implicit average over geometrical constraints. In this limit, the analytic form can be calculated from Mayer cluster functions $f_{\alpha\beta}(r) = \exp[-\beta v_{\alpha\beta}(r)] - 1$, although in practice, such procedures may be highly non-trivial[26], depending on which correlation function one chooses as well as on the complexity of the site-site potential.

Given a pair-correlation function $g_{AB}(r)$, one could also derive a description based on isotropic pairwise potentials by attempting an inversion to find the effective potential $v_g^{eff}(r)$ that reproduces $g_{AB}(r)$. If such a pairwise potential exists, then it will be unique[10-12], following arguments similar to that employed for the many-body Hamiltonians.

In the limit of zero-density, $v_g^{eff}(r)$ takes the form:

$$\lim_{\rho \to 0} \beta v_g^{eff}(r) = -\log[g_{AB}(r)] \qquad (15)$$

but in contrast to the case for a many-body Hamiltonian, this does not reduce to the bare pair potential $v_{AB}(r)$ because $g_{AB}(r)$ has a more complex dependence on the Mayer cluster integrals. It is not hard to show that, in this low density limit, $v_g^{eff}(r)$ and $v_U^{eff}(r)$ don't generally have the same analytical forms, and so suffer from similar representability problems to those we described for many-body potentials. Figure 1 foreshadows the outcome of coarse-graining the TIP4P-EW multi-site potential using the $v_g^{eff}(r)$ (Figure 1a) and $v_U^{eff}(r)$ (Figure 1b) routes, which we see generate very different functional forms.

The isothermal compressibility of a system described by a multi-site potential follows from the correlation functions[2]:
$$\rho k_B T \chi_T = 1 + 4\pi\rho \int r^2 [g_{AB}(r) - 1] dr \tag{16}$$
where the $g_{AB}(r)$ could be any one of the site-site pair-correlation functions. This expression can be used to extract thermodynamics from $v_g^{eff}(r)$. There is also clearly some freedom in choosing which correlation function is used to coarse-grain the multi-site representation to an isotropic single-site potential.

The correct statistical mechanical interpretation of a state-dependent potential like $v_g^{eff}(r)$ is highly non-trivial [13,18,20]. There is a temptation to treat such coarse-grained objects as if they really are potentials in the Hamiltonian sense, and thus exploit intuitive, but possibly misleading, analogies to physical systems described by a true pairwise potential. Although we have shown that it is straightforward to prove that representability problems are expected when multi-site potentials are coarse-grained to single site forms it is not clear how important these problems are in practice. Coarse-graining procedures can deceive, sometimes in surprising ways[13], and yet a judicious choice of coarse-graining can often lead to meaningful physical insight, even to the underlying interpretations of consistency between virial and compressibility routes to thermodynamics. To examine this dilemma in more detail, we now turn to an explicit example where we coarse-grain a multi-site water model to a single isotropic potential.

### III. Models and Methods
#### A. TIP4P-EW simulations of water
The classical non-polarizable TIP4P-EW model of water is based on the following potential energy description
$$U_{total} = U_{Coulomb} + U_{LJ} \tag{17}$$
The first term in Eq. (17) is due to Coulomb interactions
$$U_{Coulomb} = \sum_{\substack{a \in I, b \in J \\ I < J}} \frac{q_a q_b e^2}{r_{ab}} \tag{18}$$
the sum being over all charged sites $a,b$ on different molecules $I, J,$ with charges given by $q_a, q_b$. The electron charge is $e$ and $r_{ab}$ is the distance between sites. The TIP4P-EW model has three charge sites per water molecule that are placed on the two hydrogen centers and an additional site along the HOH bisector, and invokes Ewald summation to account for long-ranged electrostatics. The charges and geometries of the water molecule are given in Horn *et al* [4].

The second term in Eq. (17) is the Lennard-Jones term

$$U_{LJ} = \sum_{\substack{O \in I, O \in J \\ I < J}} v_{LJ}(r_{OO}) S(r_{OO}) + U_{LJ,tail} \tag{19}$$

where the Lennard-Jones interaction energy $v_{LJ}$:

$$v_{LJ}(r) = 4\varepsilon\left[(\sigma/r)^{12} - (\sigma/r)^{6}\right] \tag{20}.$$

acts between oxygens only, $S$ is a switching function used to avoid discontinuities due to truncation of the intermolecular potential, $U_{LJ,tail}$ is a long-range correction for the Lennard Jones interaction, and $\sigma$ and $\varepsilon$ have been optimized to the values 3.16435Å and 0.16275 kcal/mol.[4] For the TIP4P-EW model, $S$ is defined by a polynomial in $Z(r) = r^2 - R_{lower}^2$ that describes a function in the range from $Z=0$ ($r=R_{lower}$) to $Z=R_{upper}^2 - R_{lower}^2$:

$$S(Z(r)) = \begin{cases} 1 & \text{if } r \leq R_{lower} \\ 1 + AZ^3 + BZ^4 + CZ^5 & \text{if } R_{lower} < R_{upper} \\ 0 & \text{if } r > R_{upper} \end{cases} \tag{21}$$

with $A = -10/D^3$, $B = 15/D^4$, $C = -6/D^5$, and $D = R_{upper}^2 - R_{lower}^2$. This function is continuous and has continuous first and second derivatives at $r=R_{lower}$ and $r=R_{upper}$ where $R_{lower}=9.0$Å and $R_{upper}=9.5$Å.

A cubic box with edge length of ~40Å was filled with 1728 water molecules. Molecular dynamics (MD) simulations in the canonical (NVT) ensemble were performed using an in-house simulation program. The equations of motion were integrated using the velocity Verlet algorithm and a time step size of 1 femtosecond. The velocity update was done using only forces on real sites after forces on fictitious sites have been projected onto the real sites. The intra-molecular geometry ($r_{OH}$ and $\theta_{HOH}$) was constrained by applying the M_SHAKE and M_RATTLE algorithms using an absolute geometric tolerance of $10^{-10}$Å. Temperature was controlled using Nose-Hoover thermostats as described in [27]. The mass variable of the thermostats was defined by a frequency of $(0.5\text{ps})^{-1}$, except for in the diffusion runs, where the coupling was weakened to $(10.0\text{ps})^{-1}$. Coulomb interactions for the TIP4P-EW model were computed using Ewald summation. For the computation of the reciprocal space sum, 10 reciprocal space vectors in each direction were used, with a spherical cutoff for the reciprocal space sum of $n_x^2 + n_y^2 + n_z^2 \leq \sqrt{105}$. The width of the screening Gaussian was 0.35Å. The switching function in Eq. (21) (using the same settings for the switching parameters $R_{lower}$ and $R_{upper}$ as above) is also used as a molecule-based tapering function for the real-space Coulomb interaction energy in the Ewald summation.

The duration of equilibration runs was 300 picoseconds ($T > 273$ K) and 500 picoseconds ($T \leq 260$ K), using a previous high temperature production run as the start configuration at the next lower temperature. Typical production runs were between 0.6 and 2.0 nanoseconds depending on temperature. For measurement of the self-diffusion constant, five independent trajectories were run at each state point for 0.6ns each. Using the Einstein relation $6Dt = \Delta R^2(t)$, the $\Delta R^2(t) = \langle |r_1(t) - r_1(0)|^2 \rangle$ was fit using least-squares at intervals between $t$=60ps-200ps to extract the self-diffusion coefficient $D$.

**B. Inversion of pair correlation functions**

At each state point, the isotropic interaction potential $v_g^{Iso}(r)$ is constructed to have the same two body radial distribution function as the oxygen-oxygen radial distribution function, $g_{OO}(r)$, of TIP4P-EW water. In previous work we solved for the interaction potential using the Ornstein-Zernike (OZ) integral equation with the hypernetted chain (HNC) closure, given an experimental or simulated $g_{OO}(r)$ as input [5]. When the resulting isotropic OZ/HNC potential is simulated with Molecular Dynamics, it yields a reasonable estimate of $g_{OO}(r)$, but there are differences with the target TIP4P-EW structure due to approximations introduced by the HNC closure. The potential can be made more accurate by a numerical procedure which starts with the OZ/HNC solution as input to an empirical potential structure refinement (EPSR) procedure [8] in which the isotropic potential is perturbed by a constraint on the allowed $g^{Iso}(r)$

$$v_i^{Iso}(r) = v_{i-1}^{Iso}(r) + k_B T \log\left[\frac{g_{i-1}^{Iso}(r)}{g_{OO}^{TIP4P-Ew}(r)}\right] \quad (22)$$

and then iterated to self-consistency. We note that to achieve accurate results the large r portion of $v_g^{Iso}(r)$ must be handled with care The resulting numerical pair potentials are shown in Figure 2 for a number of different state points. It can be seen that this potential has two length scales, which has been shown to be a necessary, but not sufficient, condition to give anomalous properties similar to those of water [6,28]. The two lengthscales become more energetically distinct as temperature is lowered (Figure 2a), but they collapse into one lengthscale at higher density (Figure 2b).

## C. Isotropic simulations
The derived numerical isotropic pair potentials were fit using a spline interpolation and simulated using molecular dynamics in the canonical ensemble. The potentials were truncated using the switching function $S$ described above, where the cutoffs were chosen as $R_{lower}$ =15Å and $R_{upper}$=16Å for all ρ<1.7g/c. For higher ρ they were set as $R_{lower}$ =10-12Å, to maintain the highest fidelity to the $g_{OO}^{TIP4P-Ew}(r)$ out to the full box size. A timestep of 3fs was used with equilibration times as per the TIP4P-Ew simulations, and using the positions of the TIP4P-Ew oxygens from the corresponding state point as the start configurations. Nose-Hoover thermostats were again used to control temperature with the same frequencies as for the TIP4P-Ew simulations.

## IV. Results

### A. Thermodynamic properties
Thermodynamic properties can be extracted in several ways from the coarse-grained potentials we derive. One could, for example, choose the potential from a single representative state point and use it to calculate thermodynamic properties at other state points. Such an approach would be the similar in spirit to other studies where an isotropic pair potential is used to mimic certain aspects of water [5,6,28,29]. However, one could also derive a separate potential at each state point of interest and calculate thermodynamic properties with this family of potentials[24,30]. Although this does require a separate simulation with the more complex interaction at each state point in order to derive the simpler isotropic potentials, we choose to follow this route because it helps highlight the issue of representability.

By construction both models should have the same isothermal compressibility, $\kappa_T$ through Eq. (10), which we confirmed through direct comparisons. The pressure can be accessed through an integral of the compressibility over the density, and we confirm that the coarse-grained model agrees with the reference TIP4P-Ew system, as it should by construction (Figure 3a), but noting that to achieve good agreement special care should be taken to correctly sample the large $r$ limit of $g(r)$[31]. Numerical simulation and their extrapolations to large $r$ [32] still introduce some uncertainty in the large $r$ limit of the pair correlation function. Although it is then possible to construct the entire PV diagram through this route to generate the correct thermodynamics, this is less instructive as it involves an inversion at each state point and thus cannot be measured independently of the reference system.

We have also measured the virial pressure for our isotropic potential through the standard equations derived for a pairwise potential system, [2]

$$P = \rho k_B T - \frac{2\pi \rho^2}{3} \int r^3 dr g(r) \frac{dv(r)}{dr} \qquad (23)$$

and find that it does indeed deviate significantly from the TIP4P-Ew model (Figure 3(a)). The internal energy exhibits similar inaccuracies (Figure 3(b)), as expected from Figure 1, where we note the $U_{LJ,tail}$ addition to the internal energy of the TIP4P-Ew model was removed so as to only compare contributions from the N particle system. These differences in virial pressure and internal energy are considerably more pronounced than those found for a similar study of a coarse-grained polymer model [13,31], reflecting the strongly anisotropic nature of the TIP4P-Ew reference potential.

By comparing TIP4P-EW and the family of coarse-grained potentials across the entire PV diagram it is clear that these differences in magnitude between the two virial measurements are more than just a scaling factor or a constant (Figure 4). In fact, because the pair potentials change from state point to state point, they are no longer constrained by thermodynamic stability criteria. For some regions of the phase diagram the change in the coarse-grained potential $v_g(r)$ corresponding to a higher number density generates an unphysical *decrease* in pressure. As in the case of the compressibility, the virial pressure can also be derived to explicitly account for the density dependence of our family of potential [33]. We note that the Ascarelli-Harrison correction in the expression for the virial pressure, which arises for density-dependent pair potentials [33] worsens the agreement with the TIP4P-EW results (unpublished results). This echoes similar findings for coarse-graining of polymers as soft colloids (see e.g. Louis 2002, where the use of the Ascarelli-Harrison correction was strongly criticized). A more promising avenue may be to include explicit one-body potentials, as suggested by Stillinger et al [34]. The advantage for the current coarse-graining strategy is that while these one-body terms do not change the structure, they do change the thermodynamics. At present it is not clear what form they should take, but this may be a fruitful way forward that is worth exploring further.

Taking the global density dependence of the potential explicitly into account can also introduce correction terms in other quantities. For example, the expression for the compressibility can be shown to pick up extra terms [25]. For the coarse-graining strategy we follow here the basic compressibility equation (10) is by construction correct, so that extra terms will lead to deviations from the underlying TIP4P-Ew model. However, it may turn out that taking these terms into account leads to a better agreement between the compressibility and virial routes. This will be explored further in another communication.

**B. Structure**

By construction, our coarse-graining procedure generates a family of unique state-dependent single site potentials that reproduce the oxygen-oxygen pair correlations exactly. To compare the three body correlations of the TIP4P-EW and family of coarse-grained models, we measure the bond angle distribution as an integral over the full three-body correlation function, and measure the distribution of angles generated by the neighbors of each molecule within a specified radius, $R_c$.

$$b(\theta) = 8\pi^2 \rho^2 Z \int_0^{R_C} dr_{12} \int_0^{R_C} dr_{13} g^{(3)}(r_{12},r_{13},\theta) r_{12}^2 r_{13}^2 \quad (25)$$

where the triplet correlation of *eqn.2* for a spherically symmetric potential is now dependent only on the magnitude of the vectors connecting two nearest neighbors to a central particle, and the angle $\theta$ between the two vectors and $Z$ normalizes $b(\theta)$ to a probability distribution. In Figure 5 we compare the bond angle distributions for TIP4P-EW (Figure 5a) and for the isotropic family (Figure 5b). While the isotropic potentials do generate a peak at the tetrahedral angle, they show a marked increase in close-packed configurations at 60° corresponding to a defective network structure.. Even though the absolute three-body correlations are very different between the isotropic family and TIP4P-EW due to this defect structure, both models show the same trends with density, namely a loss of structural order under compression. Thus it is evident that in contrast to the virial pressure which, besides having the incorrect value at a given state point, also showed physically incorrect trends when comparing state points, the family of isotropic potentials does exhibit a structurally anomalous region, if the appropriate potential is taken at each state point. By contrast, if a single isotropic potential is taken and used at different state points, it typically does not reproduce these structural anomalies (transferability problems). We will discuss these differences in a future paper.

**C. Diffusion**
While the coarse grained particles' translation diffusion is an order of magnitude faster than the TIP4P-Ew water molecules, both models display an anomalous increase in diffusivity with compression at several isotherms. In Figure 6 we show that the TIP4P-EW (Figure 6a) and isotropic potentials (Figure 6b) reach both diffusion maxima and diffusion minima at nearly the same density along each isotherm. After the failure of these isotropic potentials to represent the correct thermodynamic and structural orientational properties of the TIP4P-Ew model, the reproduction of the dynamical anomaly is a remarkable, and perhaps unanticipated, success. Interestingly, if any one of these potentials is taken separately, and treated like a Hamiltonian system on the same density and temperature range as Figure 6, then no diffusive anomaly is seen, so that the fact that we used a different potential at each density is crucial to reproduce the correct trend.

**V. Conclusions**
In this paper we studied the coarse-graining of the multisite TIP4P-Ew model of water[4] to a representation based on a simpler isotropic pair potential. The advantage of this approach is that the properties of both the reference system and the coarse-grained system can be accurately calculated by MD simulations. By studying a series of potentials that are unique for their ability to reproduce the pair correlation function at each state point, knowing that other properties such as internal energy or virial pressure cannot be reproduced accurately, allows us to understand the relevance of this particular coarse-graining strategy.

Such careful comparisons are important because the correct interpretation of coarse-grained potentials is subtle[13,18,20]. Here we argue that their use inevitably involves some compromise[13,20]. Since it is not possible to simultaneously reproduce all the underlying properties with a single simple potential form, consumers of effective potentials need to decide which properties to focus on. For example, fitting with high accuracy to one physical property could lead to a less accurate rendering of other properties. Making an optimal choice for coarse-graining a potential may necessitate the loosening some constraints in such a way that the correct physics that one is attempting to study is properly included.

Although we have shown that representability problems can be severe for the rendering of some physical properties, we do not advocate the wholesale jettisoning of isotropic pair potentials as coarse-graining strategies for substances like water. In fact, a burgeoning community of researchers are following such strategies by developing isotropic "ramp" potentials that, notwithstanding the caveats above, nevertheless successfully mimic certain key properties of water [5,6,28,29]. These developments are important because they may help elucidate the underlying physics of water's many puzzling anomalies in terms of a competition between two lengthscales. In this regard, the systematic coarse-graining procedure we have developed here does reproduce with surprising accuracy water's dynamic anomalies. We note that the evolution of translational order is exactly captured by these coarse-grained potentials by construction, and that structural anomalies are clearly present when examined under an alternative structural metric to pure tetrahedral order. The lack of thermodynamic coherence is particularly pronounced for this coarse-graining procedure for water, making the observation of dynamic and structural anomalies within this procedure an intriguing outcome.

Representability problems are expected to apply to a much wider set of potentials than those that we describe here. For example, there are many different multi-site models for water in the literature, often designed to reproduce certain properties with more fidelity than others. See e.g.[35] for an overview of such potentials as applied to the freezing of water or [36] for a review of simpler potentials that focus on global physical properties. It is likely that some of these potentials can be improved within the constraints of their given functional forms by a more careful fitting procedure, but nevertheless they are all expected to suffer from representability problems, although these are expected to be much less severe than what we found for the thermodynamics of isotropic representations of water. It may also be useful to apply more general fitting procedures such as a Bayesian ensemble approach[37], self organizing maps[38], or adaptive resolution schemes[39], especially for complex representations with many parameters.

The work in this paper clearly shows that there is an art in knowing what needs to be preserved in the coarse-graining procedure in order to correctly render the key underlying physical processes one is trying to emulate. In the case of water's thermodynamic anomalies, higher order structural correlations may be critical, or possibly an energy scale or distribution that is not captured correctly with a coarse-graining procedure constrained to reproduce state-dependent pair correlations.

## Acknowledgements
THG gratefully acknowledges a Schlumberger Fellowship for support while on sabbatical at the University of Cambridge and financial support under DOE/BES CPIMS program.

# Appendix. Uniqueness theorem for multi-site models

While it has previously been demonstrated[10,11] that a unique pair potential will produce a unique pair distribution function, in this appendix we explicitly present the corresponding proof for an interaction-site potential describing a molecular system. Therefore we will show that a unique set of interaction-site pair potentials, $\{v_{\alpha\beta}\}$, will produce a unique set of site-site pair distribution functions, $\{g_{\alpha\beta}\}$, where $\alpha$, $\beta$ are the sites on a molecule.

We begin by defining an initial intermolecular potential, $V^0$, which can be represented as a sum of interaction-site potentials $v_{\alpha\beta}$:

$$V^0 = \sum_{\alpha,\beta}\sum_{i\neq j} v^0_{\alpha\beta}(|\mathbf{r}_{i\alpha} - \mathbf{r}_{i\beta}|) \tag{A.1}$$

which only depends on the intermolecular separations, $\mathbf{r}_{\alpha\beta} = \mathbf{r}_{i\alpha} - \mathbf{r}_{j\beta}$, between sites $\alpha$, $\beta$ on different molecules $i, j$. We can then define a new potential $V^1$ as a perturbation to $V^0$ with [11]

$$V^\lambda = V^0 + \lambda(V^1 - V^0), \tag{A.2}$$

where $\lambda$ varies between 0 and 1 and $V^1$ is defined analogously to (A.1). To demonstrate that these two sets of interaction-site potentials $\{v^1_{\alpha\beta}\}$, and $\{v^0_{\alpha\beta}\}$ will have unique sets of $\{g_{\alpha\beta}\}$, we then define the ensemble average for a multi-site potential:

$$\left\langle V^1 - V^0 \right\rangle_\lambda = \frac{\int d\mathbf{r}^N_{\alpha\beta} e^{-\beta V^\lambda}(V^1 - V^0)}{\int d\mathbf{r}^N_{\alpha\beta} e^{-\beta V^\lambda}} \tag{A.3}$$

where $d\mathbf{r}^N_{\alpha\beta}$ represents $d\mathbf{r}_{1\alpha}d\mathbf{r}_{1\beta}...d\mathbf{r}_{N\alpha}d\mathbf{r}_{N\beta}$. Differentiating (A.3) with respect to $\lambda$ gives

$$\frac{d\left\langle V^1 - V^0 \right\rangle_\lambda}{d\lambda} = -\beta\left\langle \left[(V^1 - V^0) - \left\langle V^1 - V^0 \right\rangle_\lambda\right]^2 \right\rangle_\lambda \tag{A.4}$$

such that

$$\frac{d\left\langle V^1 - V^0 \right\rangle_\lambda}{d\lambda} \leq 0. \tag{A.5}$$

Integrating the derivative between $\lambda=0$ and $\lambda=1$ gives

$$\left\langle V^1 - V^0 \right\rangle_1 = \left\langle V^1 - V^0 \right\rangle_0 + \int_0^1 d\lambda \frac{d}{d\lambda}\left\langle V^1 - V^0 \right\rangle_\lambda, \tag{A.6}$$

and combining (A.5) and (A.6) provides the relevant inequality:

$$\left\langle V^1 - V^0 \right\rangle_1 \leq \left\langle V^1 - V^0 \right\rangle_0 \tag{A.7}$$

The equality holds only for $V^1 - V^0 =$ constant. We now rewrite (A.7) substituting the interaction-site form (A.1) giving

$$\frac{\int d\mathbf{r}^N_{\alpha\beta} e^{-\beta V_1(r^N_{\alpha\beta})} \sum_{\alpha\beta}[v^1_{\alpha\beta}(r_{\alpha\beta}) - v^0_{\alpha\beta}(r_{\alpha\beta})]}{\int d\mathbf{r}^N_{\alpha\beta} e^{-\beta V_1(r^N_{\alpha\beta})}} \leq \frac{\int d\mathbf{r}^N_{\alpha\beta} e^{-\beta V_0(r^N_{\alpha\beta})} \sum_{\alpha\beta}[v^1_{\alpha\beta}(r_{\alpha\beta}) - v^0_{\alpha\beta}(r_{\alpha\beta})]}{\int d\mathbf{r}^N_{\alpha\beta} e^{-\beta V_0(r^N_{\alpha\beta})}}. \tag{A.8}$$

Using the site-site pair distribution function,

$$\rho^2 g^1_{\alpha\beta}(\mathbf{r}_{1\alpha},\mathbf{r}_{2\beta}) = \frac{N(N-1)\int d\mathbf{r}_{1\beta}d\mathbf{r}_{2\alpha}d\mathbf{r}_{3\alpha\beta}...d\mathbf{r}_{N\alpha\beta} e^{-\beta V^1(r^N_{\alpha\beta})}}{\int d\mathbf{r}^N_{\alpha\beta} e^{-\beta V^1(r^N_{\alpha\beta})}}, \tag{A.9}$$

and noting these distributions again depend only on the separation, $\mathbf{r}_{\alpha\beta} = \mathbf{r}_{i\alpha} - \mathbf{r}_{j\beta}$, the inequality becomes

$$\sum_{\alpha,\beta} \int d\mathbf{r}_{\alpha\beta} \left[ v^1_{\alpha\beta}(r_{\alpha\beta}) - v^0_{\alpha\beta}(r_{\alpha\beta}) \right]\left[ g^1_{\alpha\beta}(\mathbf{r}_{\alpha\beta}) - g^0_{\alpha\beta}(\mathbf{r}_{\alpha\beta}) \right] \leq 0. \qquad (A.10)$$

By choosing $\{v^1_{\alpha\beta}\}$ to be unique from $\{v^0_{\alpha\beta}\}$ i.e. they differ by more than a constant, then requires the left hand side of (A.9) to be nonzero and that therefore the set of $\{g^1_{\alpha\beta}\}$ must differ from the set of $\{g^0_{\alpha\beta}\}$.

## Figure Captions

**Figure 1.** *The effective potential through the internal energy (stars) and pair correlation function (circles) coarse-graining procedures for* (a) T=310K and ρ=0.9g/cc and (b) T=310K and ρ=1.29g/cc, illustrating the difference between the unique potential $v_g(r)$ generated through the pair correlation function route and $v_U(r)$ generated through the internal energy route. Clearly the former potential will not reproduce the internal energy through the standard formula (3).

**Figure 2.** *Isotropic potentials derived through the pair correlation route;* (a) The family of isotropic potentials derived from the TIP4P-EW $g_{OO}(r)$ at P=1atm and for temperatures 235.5K, 248K, 260.5K, 273K, 285.5K, 298K and 310.5K. (b) The family of isotropic potentials derived from TIP4P-EW gOO(r) at T=235.5K and for densities 0.9g/cc, …1.29g/cc.

**Figure 3.** (a) The dimensionless compressibility factor $Z=\langle P/\rangle$ vs T along the P=1atm isobar for *TIP4P-EW and the family of isotropic potentials.* The pressure for the isotropic potentials is measured through both the compressibility route, which correctly reproduces the TIP4P-EW result by construction, and through the virial route, which does not. Z=1 represents the ideal gas. Inset demonstrates that agreement is not perfect due to errors introduced by numerical integration of compressibility equation, and extrapolation of g(r) to large *r.* (b) The internal energy $\langle U/N$ for the TIP4P-EW and the family of isotropic potentials. The simulated values are shown with the LJtail correction removed from the TIP4P-Ew values.

**Figure 4.** *Pressure-Volume phase diagram for TIP4P-EW and the family of isotropic potentials along four isotherms 310.5 (stars), 285.5 (triangles), 260.5 (squares), 235.5 (circles).* (a) TIP4P-Ew: The density anomaly occurs in regions where lower isotherms cross above a higher T isotherm. (b) Isotropic potentials: The pressure is the average virial pressure from each simulated point, and the lines act only as a guide.

**Figure 5:** *Bond angle distributions at T=235.5K as a function of density for* (a) TIP4P-Ew and (b) Isotropic potentials. $R_C$ was chosen as 3.4A.

**Figure 6.** *Translational diffusion constants vs density along four isotherms 310.5 (stars), 285.5 (triangles), 260.5 (squares), 235.5 (circles). Lines are fifth order polynomial fits to data points* (a) TIP4P-Ew and (b) Isotropic potentials.

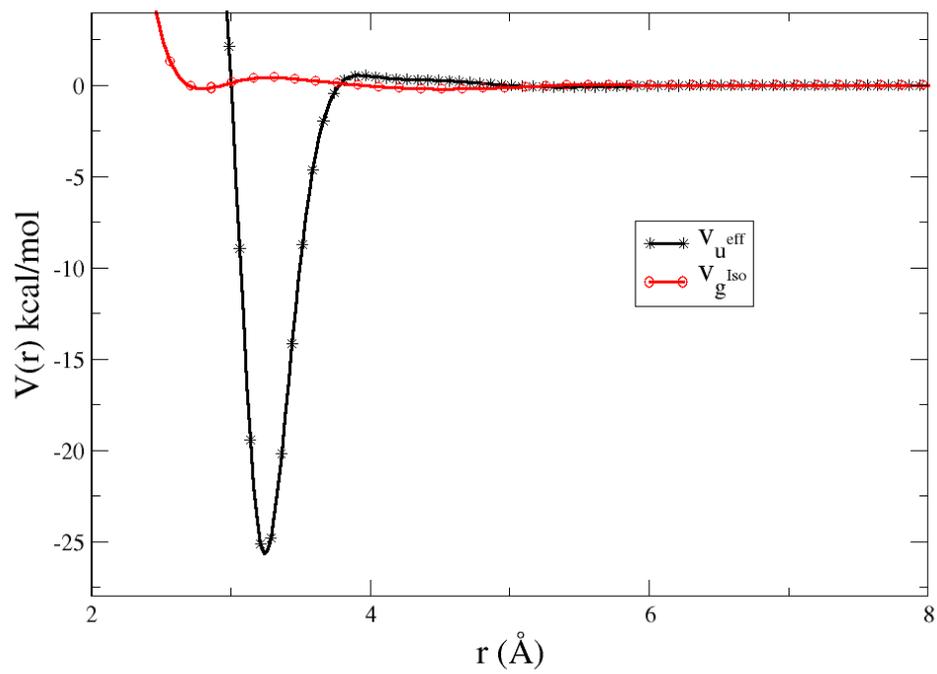

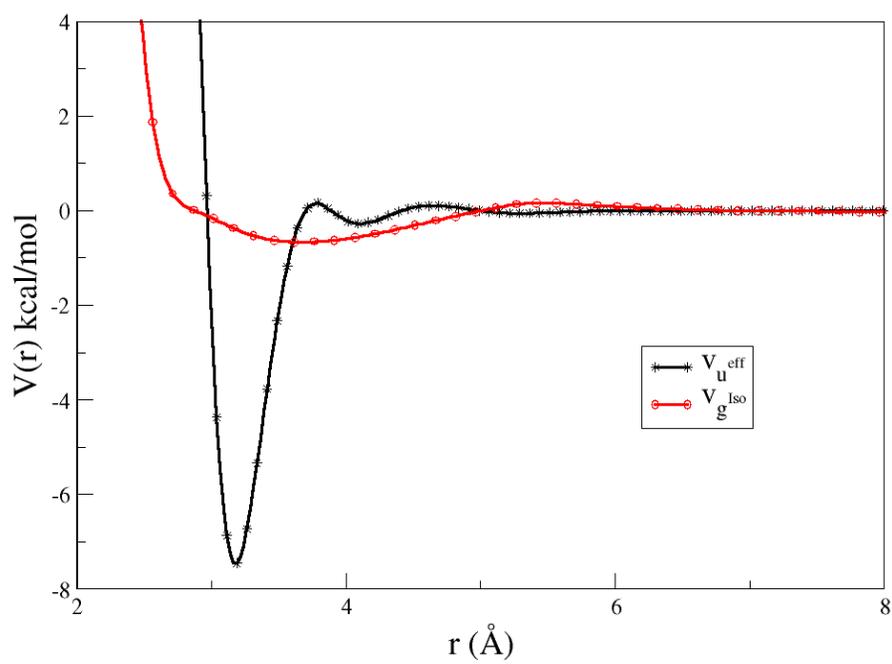

**Figure 1.** Johnson et al.

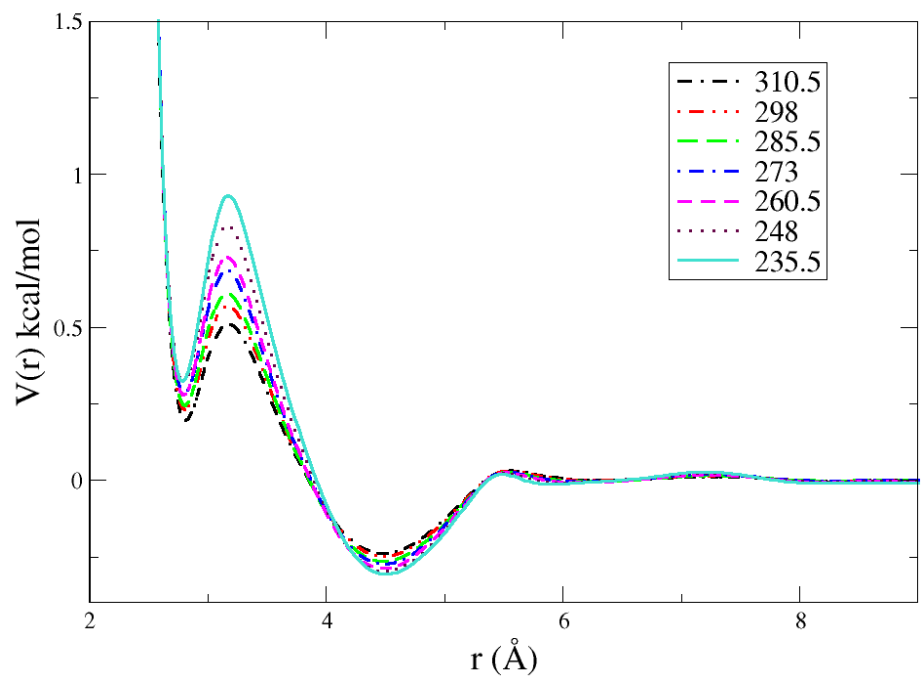

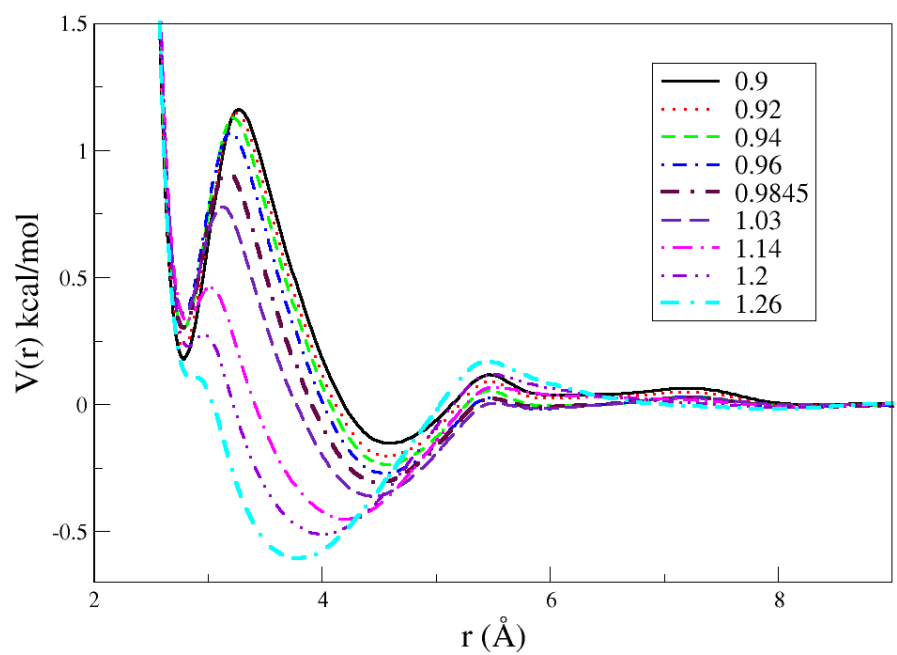

**Figure 2.** Johnson et al.

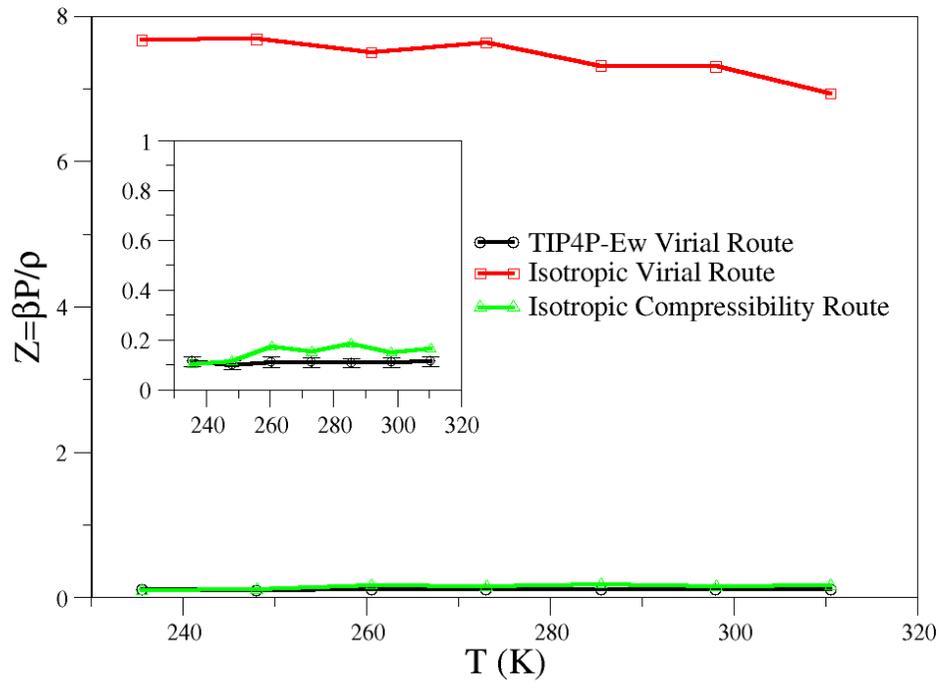

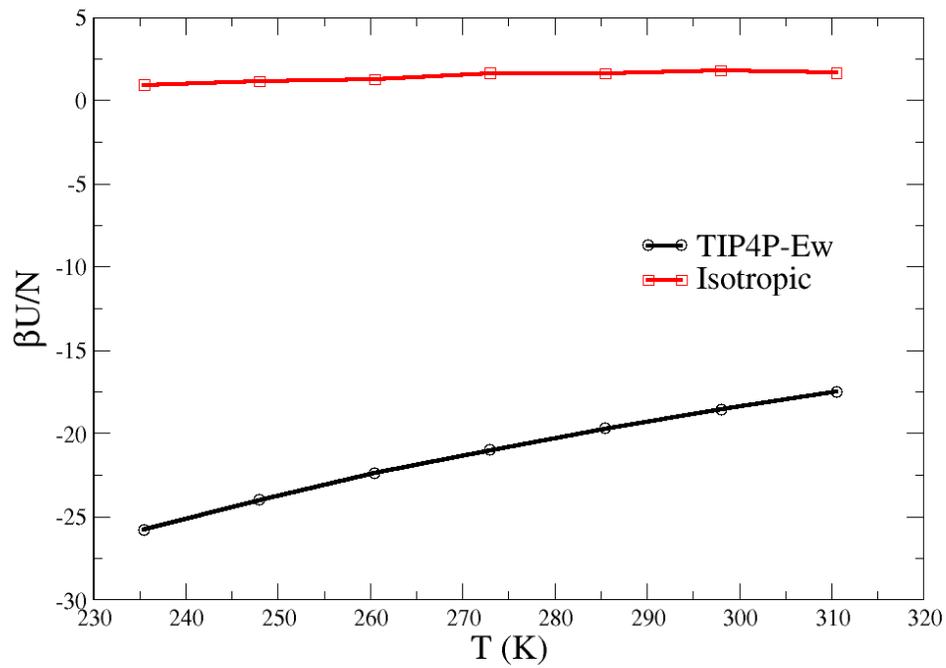

**Figure 3.** Johnson et al.

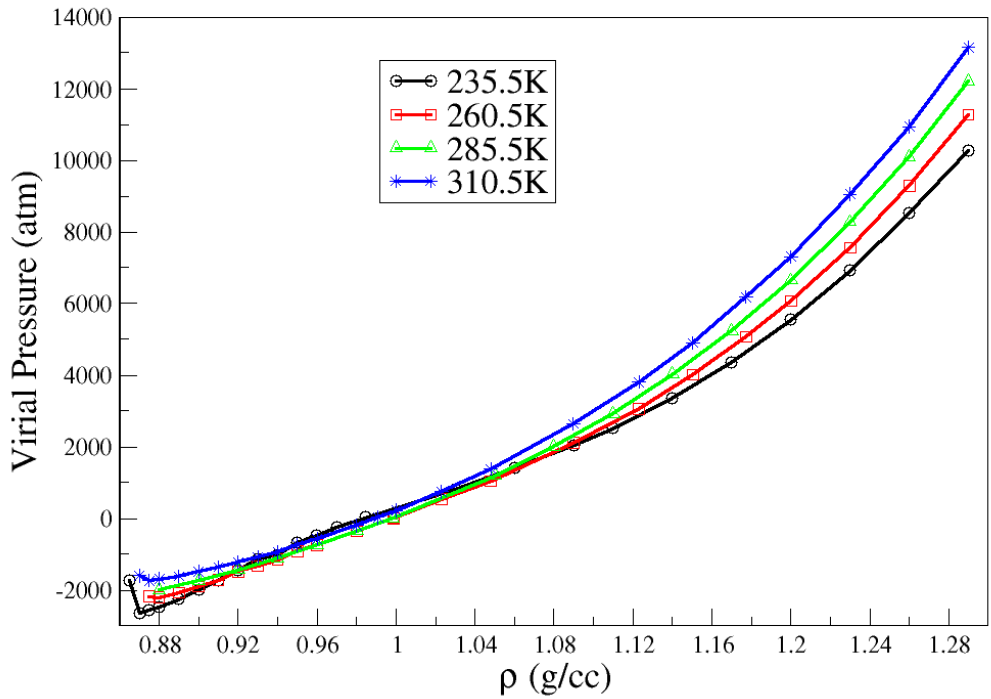

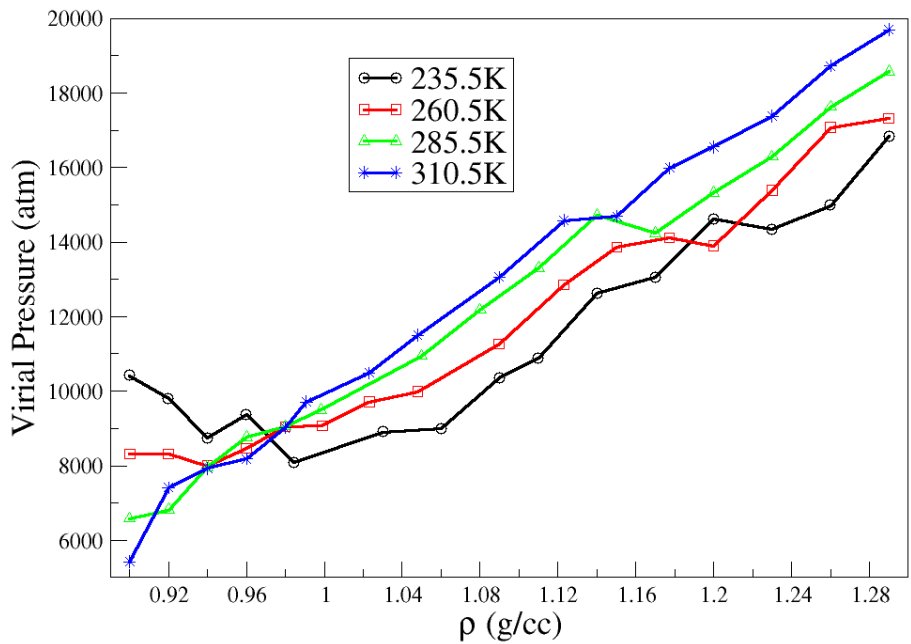

**Figure 4.** Johnson et al.

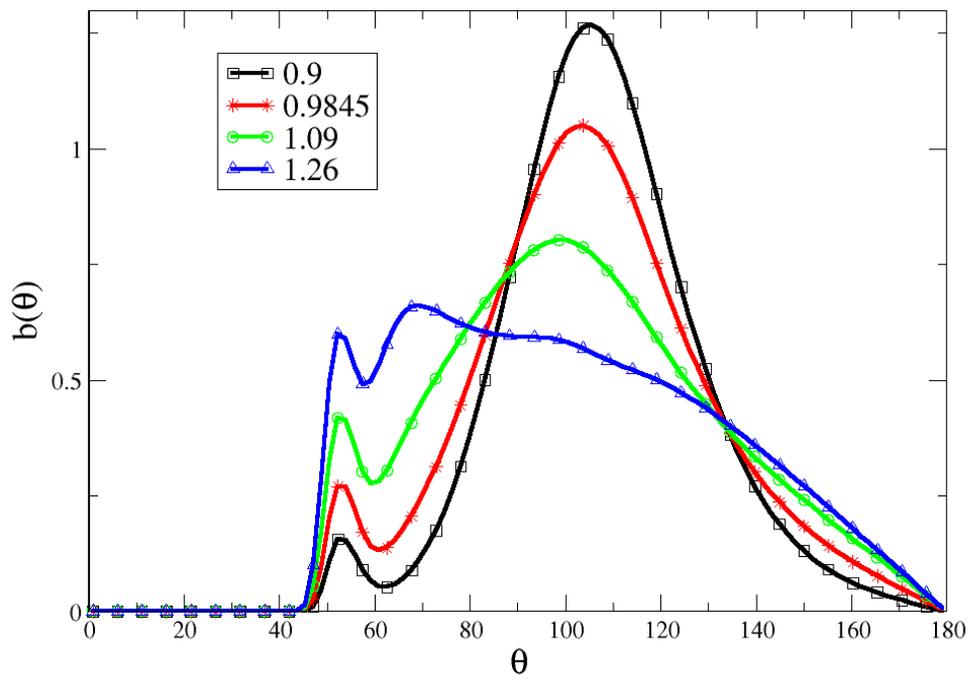

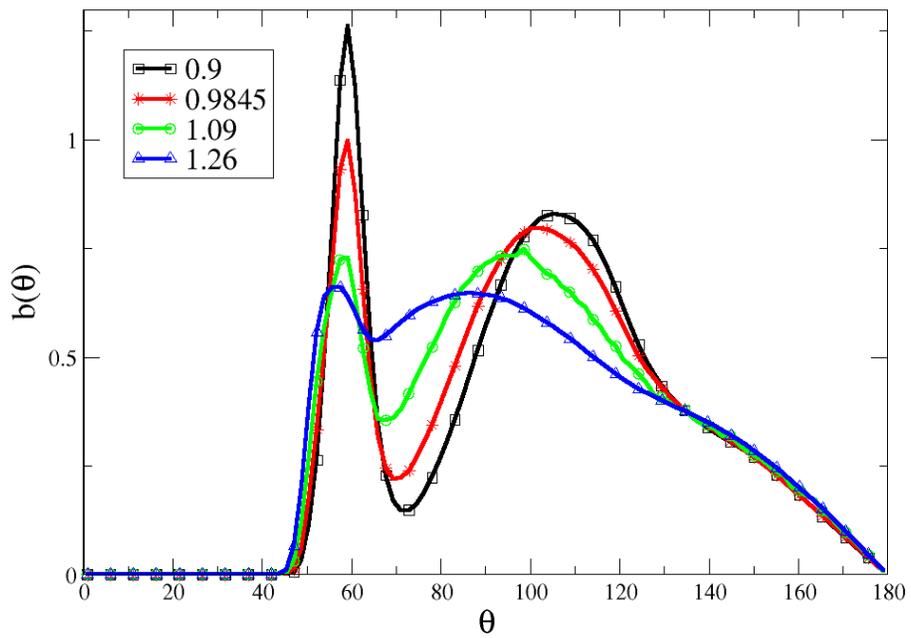

**Figure 5.** Johnson et al.

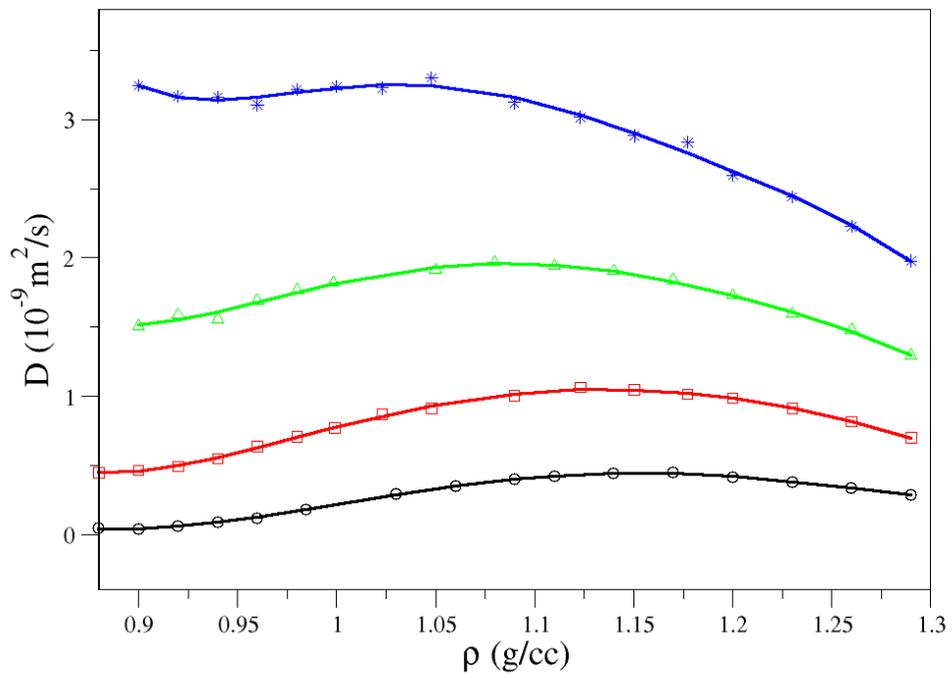

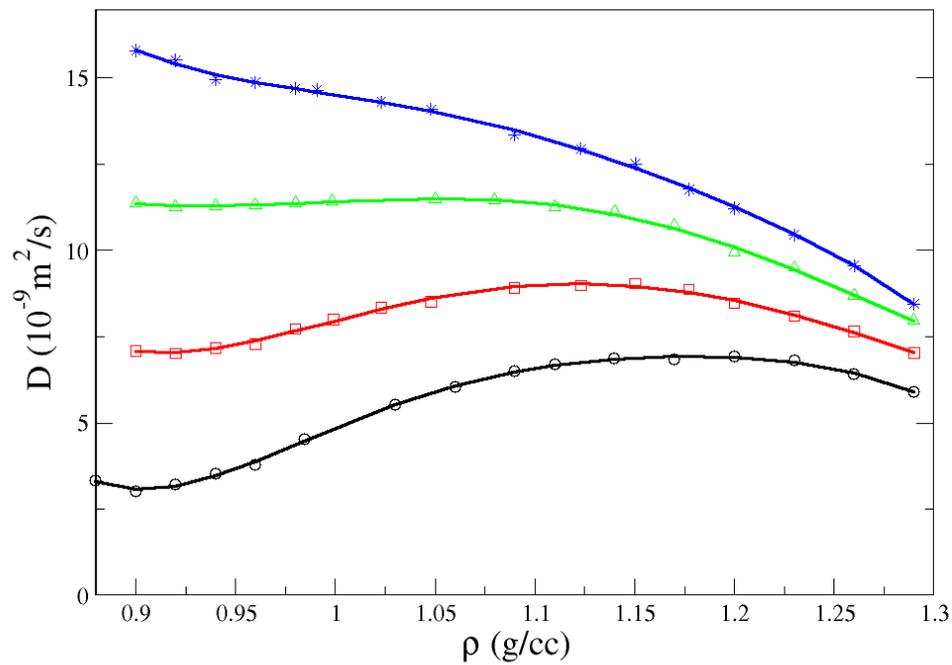

**Figure 6.** Johnson et al.